\documentclass[
 reprint,
 amsmath,amssymb,
 aip,
 apl,
 floatfix,
]{revtex4-2}

\usepackage{graphicx}
\usepackage{bm}
\usepackage[utf8]{inputenc}
\usepackage[T1]{fontenc}
\usepackage{hyperref}
\usepackage[all]{hypcap}
\usepackage[capitalise]{cleveref}

\begin{document}

\title{First principles design of Ohmic spin diodes based on quaternary Heusler compounds}

\author{T. Aull$^{1}$}\email{thorsten.aull@physik.uni-halle.de}
\author{E. \c{S}a\c{s}{\i}o\u{g}lu$^{1}$}
\author{I. Mertig$^{1,2}$}

\affiliation{$^{1}$Institute of Physics, Martin Luther University Halle-Wittenberg, D-06120 Halle (Saale), Germany \\
$^{2}$Max Planck Institute of Microstructure Physics, Weinberg 2, D-06120 Halle (Saale), Germany}

\begin{abstract}

The Ohmic spin diode (OSD) is a recent concept in spintronics, which is based on half-metallic magnets 
(HMMs) and spin-gapless semiconductors (SGSs). Quaternary Heusler compounds offer a unique platform to 
realize the OSD for room temperature applications as these materials possess very high Curie temperatures 
as well as  half-metallic and spin-gapless semiconducting behavior within the same family. Using 
state-of-the-art first-principles calculations combined with the non-equilibrium Green's function method we 
design four different OSDs based on half-metallic and spin-gapless semiconducting quaternary Heusler compounds. All four OSDs exhibit linear 
current-voltage ($I-V$) characteristics with zero threshold voltage $V_T$. We show that these OSDs 
possess a small leakage current, which stems from the overlap of the conduction and valence band edges of 
opposite spin channels around the  Fermi level in the SGS electrodes. The obtained on/off current ratios vary 
between $30$ and $10^5$. Our results can pave the way for the experimental fabrication of the OSDs within the
family of ordered quaternary Heusler compounds.

\end{abstract}

\maketitle

Spintronics is a rapidly emerging field in current nanoelectronics. Due to their 
diverse and tunable electronic and magnetic properties, Heusler compounds received 
great interest for potential applications in spintronics. Especially, within the last two 
decades half metallic Heusler compounds with 100\% spin polarization of the conduction 
electrons at the Fermi energy~\cite{de1983new} have been extensively studied; both, 
theoretically and experimentally, for memory and sensor applications. Besides half metallicity
in ordinary X$_2$YZ-type Heusler compounds, several quaternary Heuslers with chemical formula 
\textit{XX'YZ}, with  $X$, $X'$ and $Y$ are transition-metal atoms and $Z$ is an $sp$ element, 
have been theoretically predicted to exhibit spin-gapless semiconducting behavior and 
some of them have been experimentally synthesized \cite{ozdougan2013slater,galanakis2016spin,gao2019high,ouardi2013realization}. 
Spin-gapless semiconductors (SGSs) possess a unique electronic structure, in which conduction- 
and valence-band edges of opposite spins touch at the Fermi level~\cite{wang2008proposal} 
and thus SGS behavior leads to unique functionalities and device concepts.

\begin{figure}[!b]
\centering
\includegraphics[width=8.5cm]{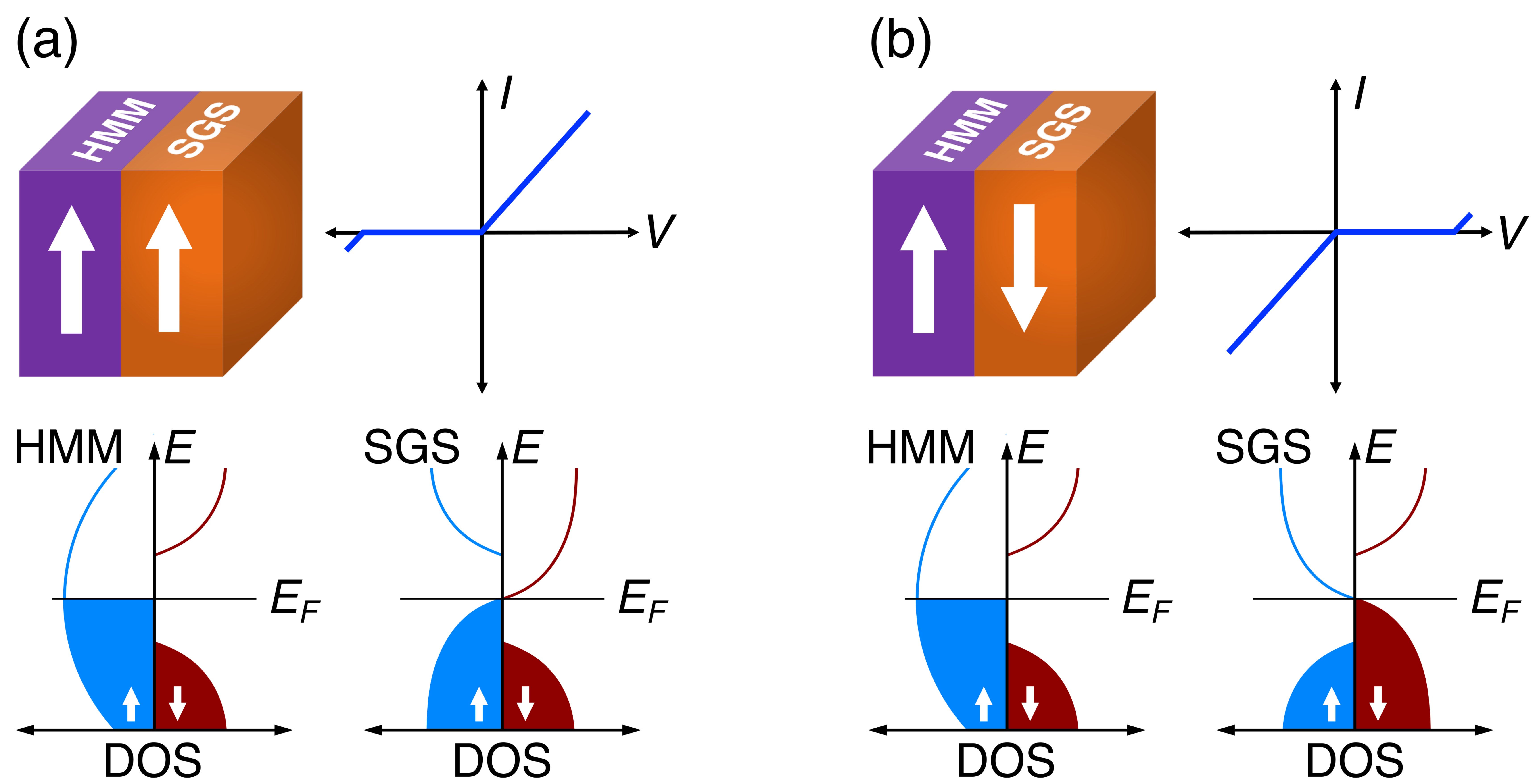}
\caption{(Color online) (a) Upper part: Schematic representation of the Ohmic spin diode and 
corresponding current-voltage ($I-V$) characteristics. Arrows show the magnetization direction 
of the  HMM and SGS electrodes (ferromagnetic OSD). Lower part: Schematic representation of the 
density of states for a HMM and SGS. (b) The same as (a) for anti-ferromagnetic coupling of the 
HMM and SGS electrodes (antiferromagnetic OSD).}
    \label{fig1}
\end{figure}

\begin{table*}
\caption{\label{tab:StructureParam}
Material composition of the considered OSDs, coupling of the electrodes, lattice constants $a_0$, $c/a$ ratio, 
sublattice and total magnetic moments, work function ($\Phi$), the magnetic anisotropy energy (MAE), Curie 
temperatures $T_C$ of the cubic phase and the electronic ground state. The $T_C$ values are taken from 
Ref.~\onlinecite{aull2019ab}.}
\begin{ruledtabular}
\begin{tabular}{@{}*{1}l*{1}c*{1}l*{10}{c}@{}}
& & & $a_0$ & $c/a$ & m$_X$ & m$_{X'}$ & m$_Y$ & m$_\text{total}$ & $\Phi$ & MAE\footnotemark[1] & $T_C$ & Ground \\
 HMM-SGS junction & Coupling & Compound & ({\AA}) & & ($\mu_B$) & ($\mu_B$) & ($\mu_B$) & ($\mu_B$) & (eV) & ($\mu$eV/at.) & (K) & state \\
\hline
MnVTiAl$-$FeVTaAl & $\uparrow \uparrow$ & FeVTaAl  & 6.10 & 1.00 & 0.85  & 2.38 & -0.19 & 3.00 & 3.75 & 0.63   & 681 & SGS \\
                  & & MnVTiAl  & 6.10 & 1.01 & -2.42 & 2.61 & 0.86  & 1.00 & 3.59 & 11.94  & 963 & HMM \\
FeVHfAl$-$FeVTiSi & $\uparrow \uparrow$ & FeVTiSi  & 5.91 & 1.00 & 0.57  & 2.33 & 0.10  & 3.00 & 3.52 & -0.94  & 464 & SGS \\
                  & & FeVHfAl  & 5.91 & 1.12 & -0.15 & 2.06 & 0.10  & 2.00 & 4.10 & 117.44 & 742 & HMM \\
FeVHfAl$-$FeVNbAl & $\uparrow \uparrow$ & FeVNbAl  & 6.11 & 1.00 & 0.81  & 2.32 & -0.11 & 3.00 & 3.72 & 0.25   & 693 & SGS \\  
                  & & FeVHfAl  & 6.11 & 1.04 & -0.68 & 2.41 & 0.29  & 2.00 & 3.45 & 62.44  & 742 & HMM \\
Co$_2$MnSi$-$FeVTaAl & $\uparrow \downarrow$ & FeVTaAl  & 6.10 & 1.00 & 0.79  & 2.32 & -0.11 & 3.00 & 3.75 & 0.63   & 681 & SGS \\
                     &  & CoCoMnSi & 6.10 & 0.86 & 1.01  & 1.01 & 3.18  & 5.00 & 3.83 & 57.50  & 920 & HMM
\end{tabular}
\end{ruledtabular}
\footnotetext[1]{Out-of-plane magnetisation is marked as negative MAE}
\end{table*}

Half-metallic Heusler compounds have been considered as ideal electrode materials in 
magnetic tunnel junctions for spin-transfer torque magnetic memory applications due 
to their very high Curie temperatures. The use of Co-based Heusler compounds in magnetic 
tunnel junctions made the experimental observation of high tunnel magnetoresistance (TMR) effects possible.~\cite{kammerer2004co2mnsi,sakuraba2006giant,wang2009giant,xu2013new,faleev2017heusler}
However, magnetic tunnel junctions constructed with half metals do not present any rectification
(or diode effect) for logic operations. Lately logic functionality in magnetic tunnel
junctions is achieved by replacing one of the electrodes by a SGS material.  In 
Ref.~\onlinecite{sasioglu2019proposal}, based on HMMs and SGSs, a reconfigurable magnetic 
tunnel diode and transistor has been proposed. This concept combines logic and memory 
on the diode and transistor level. Moreover, in a recent publication the present authors 
proposed another device concept based on HMMs and SGSs, which is the so-called Ohmic spin diode 
(OSD)~\cite{SpinDiode}. It has been computationally demonstrated that the OSD comprising 
two-dimensional half-metallic Fe/MoS$_2$ and spin-gapless semiconducting VS$_2$ exhibits 
linear current-voltage ($I-V$) characteristics with zero threshold voltage $V_T$. OSDs 
have a much higher current drive capability and low resistance, which is advantageous 
compared to conventional semiconductor $p-n$ junction diodes and metal-semiconductor 
Schottky diodes.

The aim of the present Letter is a computational design of OSDs based on quaternary
Heusler compounds for room temperature applications. Heusler compounds offer a unique 
platform to realize the OSD as these materials possess very high Curie temperatures
(much above room temperature) as well as half-metallic and spin-gapless semiconducting 
behavior within the same family. To this end, the selection of the SGS and HMM electrode 
materials from the  Heusler family for the design of OSDs is based on our recent study 
in Ref.~\onlinecite{aull2019ab}, where we focus on Curie temperatures, spin-gaps, formation 
energy, Hull distance for a large number of quaternary Heusler compounds. Among the considered
materials  three SGSs (FeVNbAl, FeVTaAl, and FeVTiSi) turn out to be promising for device 
applications. As for the  half-metallic Heusler compounds, we have a large variety of
choice, but we stick to the quaternary ones (MnVTiAl, FeVHfAl) with similar lattice constants 
to the SGSs in order to ensure a coherent growth on top of each other. Additionally, we also 
consider the well-known half-metallic Co$_2$MnSi system among the ordinary full Heusler compounds 
as an electrode material.

Our first-principles design of the OSDs is based on the density functional theory (DFT) using the  
\textsc{QuantumATK} package (version P-2019.12)~\cite{QuantumATK,smidstrup2019an}.
As exchange-correlation functional we chose the Perdew-Burke-Ernzerhof (PBE) parametrization~\cite{perdew1996generalized} 
combined with norm-conserving PseudoDojo pseudopotentials~\cite{QuantumATKPseudoDojo} and linear 
combinations of atomic orbitals (LCAO) as basis-set. As $\mathbf{k}$-point grid for the ground state 
properties we use a $15 \times 15 \times 15$ Monkhorst-Pack grid and a density mesh cutoff of 120 
Hartree. For the transport calculations we combined DFT with the nonequilibrium Green's function method 
(NEGF). We use a $15 \times 15 \times 160$ $\mathbf{k}$-point mesh in self-consistent DFT-NEGF calculations.
The $I-V$ characteristics were calculated within the Landauer approach~\cite{Landauer-Buettiker}, 
where $I(V) = e/h \sum_{\sigma}\int \, T^{\sigma}(E,V)\left[f_{L}(E,V)-f_{R}(E,V)\right] \mathrm{d}E $. 
$V$ stands for the bias voltage and the transmission coefficient $T^\sigma (E,V)$ depends additionally 
on spin $\sigma$ of an electron and energy $E$. Furthermore, $f_L(E,V)$ and $f_R(E,V)$ denote 
the Fermi-Dirac distribution of the left and right electrode, respectively. For the calculation of 
$T^\sigma (E,V)$ we chose a $25 \times 25$ $\mathbf{k}$-point grid.

\begin{figure*}
\centering
\includegraphics[width=17cm]{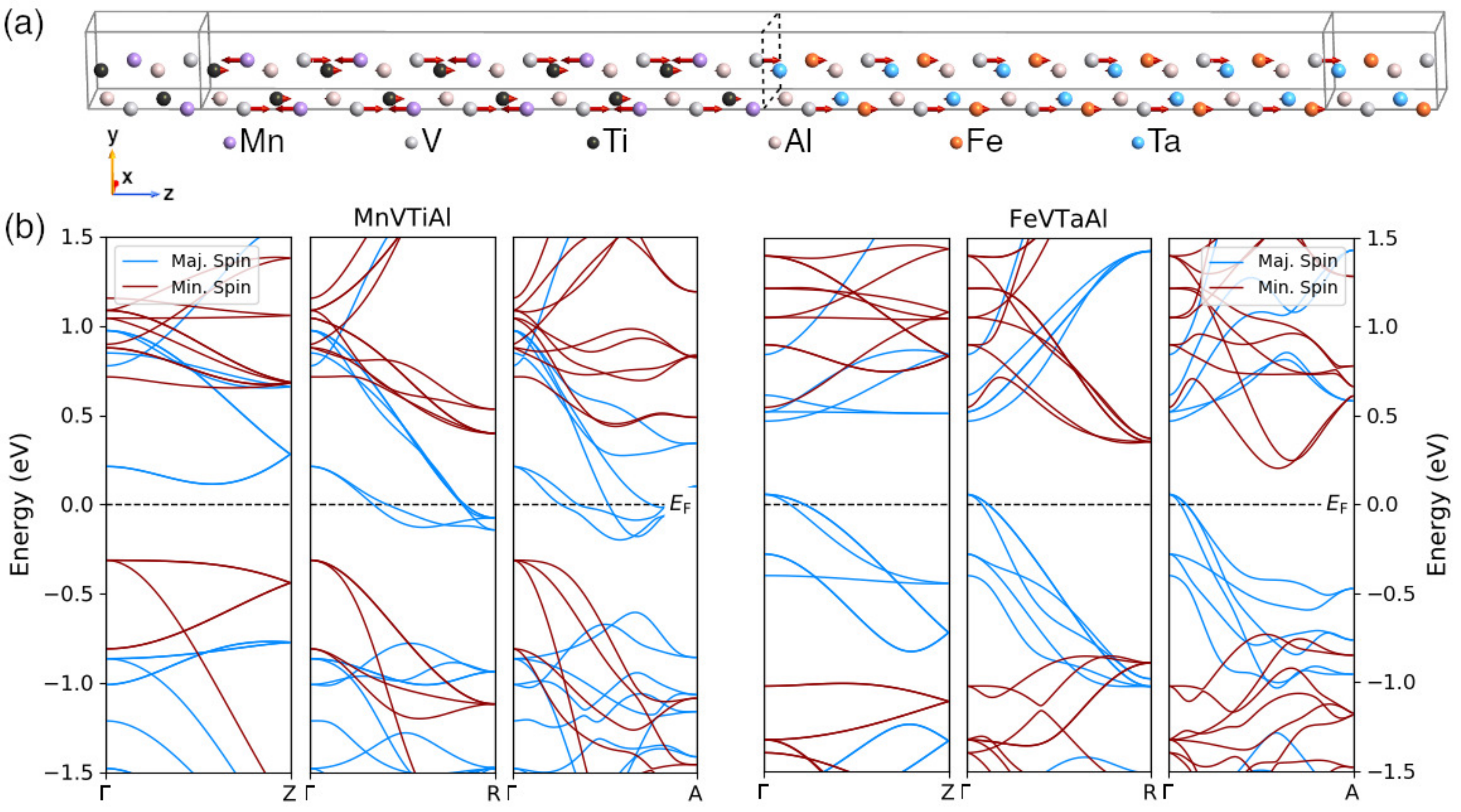}
\caption{(Color online) (a) The atomic structure of the MnVTiAl$-$FeVTaAl Ohmic spin diode.
The system is periodic in $x$- and $y$-direction in the plane orthogonal to the $z$-direction, 
which is the transport direction. The red arrows illustrate the 
direction as well as the magnitude of the magnetic moments within the scattering region. Small 
magnetic moments are overlayed by the atomic radii. The black dashed box denotes the interface. 
(b) The calculated spin-resolved bulk electronic band structure along the device stack direction, [001], for MnVTiAl (left panel) and FeVTaAl (right panel). For both compounds the 
Fermi level is set to zero (dashed black line).}
\label{fig2}
\end{figure*}

In \cref{fig1} we present schematically the structure of the OSD and the corresponding 
$I-V$ characteristics. The concept of the OSD has been extensively discussed in Ref.~\onlinecite{SpinDiode}
and thus here we will only give a short overview of the device. The OSD consists of HMM and SGS 
materials and possesses linear $I-V$ characteristics. The schematic DOS of these materials are also shown 
in \cref{fig1}. Depending on the choice of the junction materials, the HMM and SGS electrodes can couple 
ferromagnetically (ferromagnetic OSD) or antiferromagnetically (antiferromagnetic OSD) at the interface 
giving rise to corresponding $I-V$ curves shown in \cref{fig1}. The operation principle of the OSD 
relies on the unique spin-dependent transport properties of HMMs and SGSs as discussed in Ref.~\onlinecite{SpinDiode} 
in detail.

In the proposal of the OSD~\cite{SpinDiode}, the proof of principle was 
demonstrated by using two dimensional transition-metal dichalcogenides VS$_2$ (SGS) 
and Fe/MoS$_2$ (HMM) as electrode materials. Since VS$_2$ possesses an estimated
Curie temperature of 138\,K~\cite{fuh2016newtype}, it is not suitable for room
temperature applications. For more realistic devices we now consider six Heusler compounds as 
mentioned before and construct four different OSDs: i) FeVHfAl$-$ FeVTiSi, ii) FeVHfAl$-$FeVNbAl, 
iii) MnVTiAl$-$FeVTaAl, and iv) Co$_2$MnSi$-$FeVTaAl. All six Heusler compounds possess extremely 
high Curie temperatures as presented in \cref{tab:StructureParam}. For the construction of the OSD 
we assume the situation where one material needs to be grown on top of the other one. Thus, in our simulations we
take one electrode (SGS) in the cubic structure and relax the second electrode material (HMM) with respect to the in-plane lattice parameter of the first one. Therefore, in \cref{tab:StructureParam} we include the $c/a$ ratios for the half metallic electrode materials which takes the tetragonal structure.
In \cref{tab:StructureParam} 
we present also the obtained magnetic moments, magnetic anisotropy energies (MAEs), and work functions. As expected, tetragonal 
distortion results in a significant change in the magnetic anisotropy energy of HMMs, which 
is at least two orders of magnitude larger than the SGSs, being in good agreement with the 
literature~\cite{kuroda2020first}.

\begin{figure*}
\includegraphics[width=0.9\textwidth]{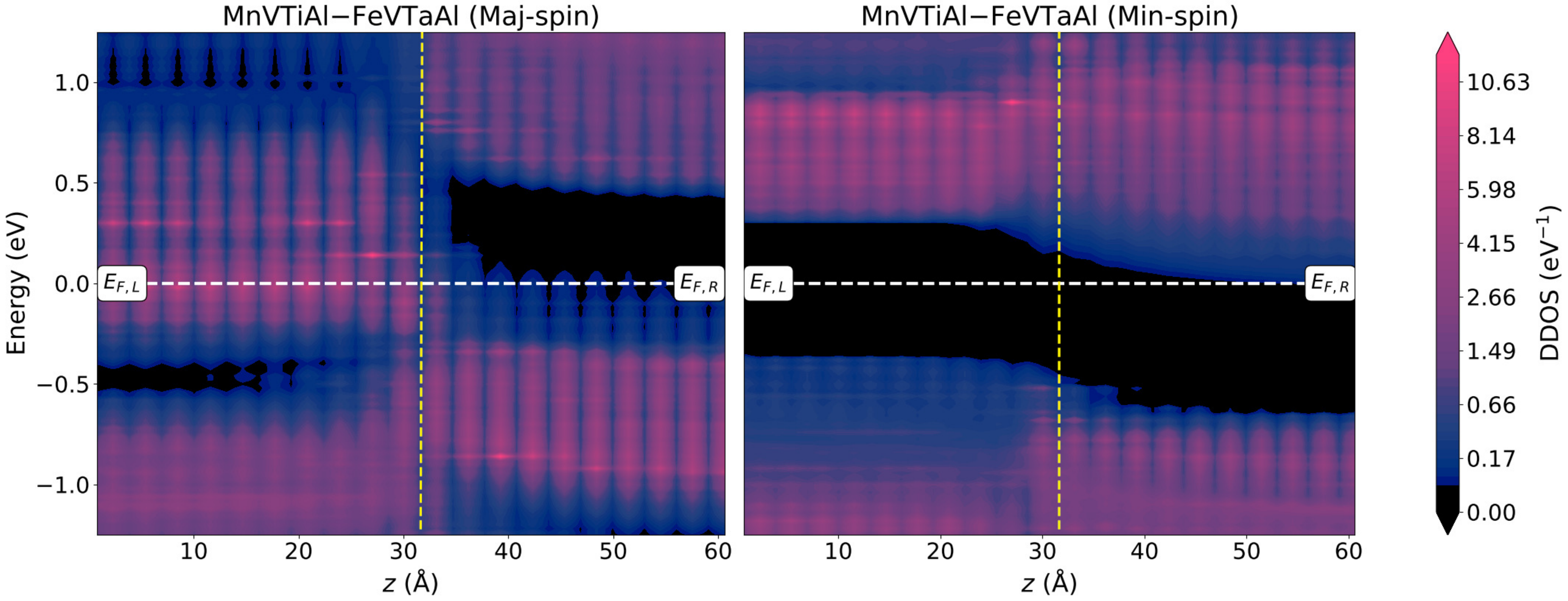}
\caption{(Color online) Projected device density of states (DDOS) at zero bias (equilibrium) for the 
majority (left panel) and minority spin channel (right panel) of the MnVTiAl$-$FeVTaAl OSD (the atomic 
structure is given in \cref{fig2}). The white dashed lines display the Fermi level while the vertical 
yellow dashed lines denote the interface.}
\label{fig3}
\end{figure*}

We now focus on the first OSD and discuss its structural, electronic and magnetic properties.
Fig.~\hyperref[fig2]{2\,(a)} illustrates the atomic structure of the OSD based on half-metallic MnVTiAl 
(left electrode) and spin-gapless semiconducting FeVTaAl (right electrode) quaternary Heusler 
compounds. We use a minimal tetragonal unit cell along the [001] direction containing 8 atoms. For each electrode this cells was repeated 5 times and defines the screening region.
The length (screening region) of the device is around 62\,{\AA}.
In the other three OSDs, the considered screening region lies in between 61{\AA} and 63{\AA}, depending 
on the materials. As the strength of the spin-orbit coupling (SOC) is very weak in these materials we, 
neglect the SOC in transport calculations and thus we chose the $z$-direction as the transport direction and also adjusted the alignment of the magnetic moments to the $z$-axis. The red arrows and their size in Fig.~\hyperref[fig2]{2\,(a)} 
represent the direction and magnitude of the atomic magnetic moments in the junction materials. 
In this OSD, the HMM and SGS electrodes couple ferromagnetically at the interface, i.e., the energy 
difference between ferromagnetic and antiferromagnetic coupling is about 400\,meV.

Looking at the magnetic moments at the interface region in \cref{fig2}, we notice that the size of the arrows
deviate from their bulk behavior, i.e, far from the interface. Especially the magnetic moment of the Mn atom in MnVTiAl 
at the interface decreases from 2.42\,$\mu_B$ to 0.07\,$\mu_B$. This is due to the fact, that at the interface region
MnVTaAl is formed which also presents HMM behavior with bulk magnetic moments of $m_{Mn}=0.08\,\mu_B$, 
$m_{V}=1.97\,\mu_B$, $m_{Ta} = -0.04\,\mu_B$ and $m_{Al} = -0.01\,\mu_B$. Therefore, the half metallic character 
of the MnVTiAl compound is retained at the interface. However, the FeVTaAl compound loses its spin-gapless 
semiconducting nature near the interface region as it will be discussed later in detail.

Next, we would like to discuss the electronic properties of the MnVTiAl$-$FeVTaAl junction at equilibrium, 
i.e., at zero bias. The bulk band structure along the transport direction of the junction materials is shown in Fig.~\hyperref[fig2]{2\,(b)}.
MnVTiAl is a HMM with a band gap of around 650\,meV in the minority-spin channel while FeVTaAl shows SGS properties. Note that the spin-gapless semiconducting behavior along the chosen directions is not well seen and for a detailed discussion the reader is referred to Refs.~\onlinecite{gao2019high} and \onlinecite{aull2019ab}.
Nikolaev \textit{et al.}~\cite{nikolaev2009all} and Bai \textit{et al.}~\cite{bai2013high} provided a detailed 
discussion about the importance of band matching for the transport properties of giant magnetoresistance (GMR) spintronic devices.
In our case, as seen in Fig.~\hyperref[fig2]{2\,(b)}, there is a good band matching for the majority spin states near the 
Fermi level close to the $\Gamma$-point between the electrode materials, especially along the $\Gamma$-R and $\Gamma$-A directions. 
Note that a good band matching suppresses the electron back scattering at the interface and ensures a smooth 
propagation of majority spin electrons from the FeVTaAl electrode to the MnVTiAl electrode.
In \cref{fig3} we present the device density of states (DDOS) of the 
MnVTiAl$-$FeVTaAl junction. As mentioned above the half metallicity of MnVTiAl is preserved at the interface
which can be seen in the majority-spin and minority-spin channel DDOS presented in \cref{fig3}. 
However, the SGS character of FeVTaAl is lost near the interface region
due to the charge transfer 
from the half-metallic electrode to the SGS (see the left panel of \cref{fig3}). 
This charge transfer stems from the work function difference of 160\,meV 
between the two electrode materials. As MnVTiAl has the lower work function, electrons flow from the majority-spin
channel of MnVTiAl to the majority spin-channel of FeVTaAl.
Once the charge redistribution reaches equilibrium, 
MnVTiAl will be positively charged near the interface region, whereas 
FeVTaAl will be negatively charged and hence an electric dipole will be induced. 
This dipole influences the electronic and magnetic properties of both materials.
Since the magnetic moment of the Mn atom in the MnVTiAl electrode has already been discussed above, we will 
briefly comment on the magnetic moment of the V atoms near the interface region.The magnetic moment of the V atoms in MnVTiAl is reduced to 2.25$\mu_B$ close to the interface and recovers to the bulk value within two unit cells. 
On the other hand, the variation of magnetic moments 
in the FeVTaAl electrode near the interface region is around 0.1$\mu_B$. Thus, the affected region by the charge transfer is restricted to five atomic layers for the majority-spin channel. The minority-spin channel is not substantially affected for both junction materials.

\begin{figure*}
\centering
\includegraphics[width=0.9\textwidth]{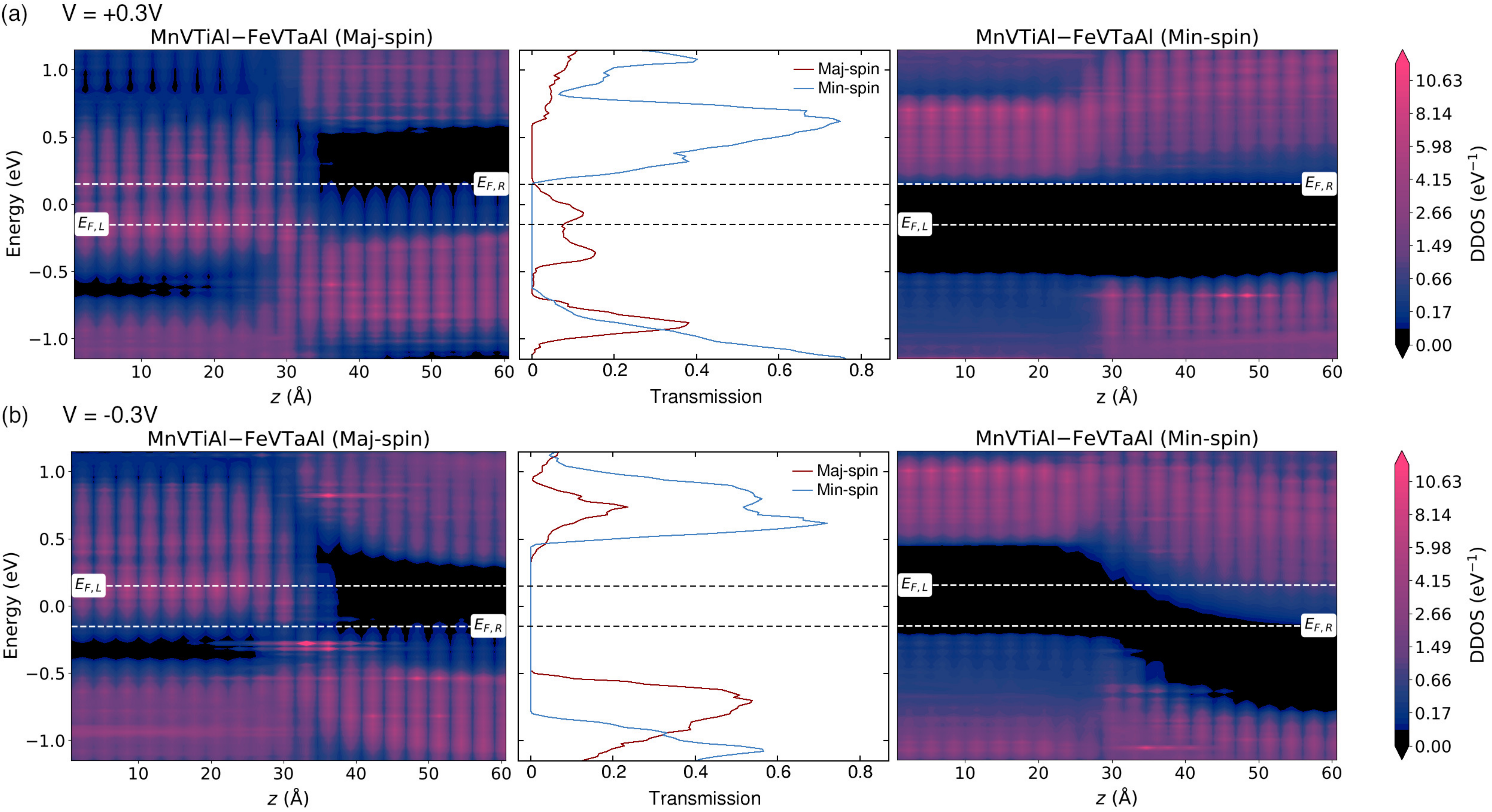}
\caption{(Color online) (a) Projected local device density of states (DDOS) for the majority (left panel) 
and minority (right panel) spin channel in MnVTiAl$-$FeVTaAl OSD (the atomic structure of 
an OSD is provided in \cref{fig2}) for a bias voltage of $V=0.3$\,V. In the middle panel we present the 
calculated transmission spectrum for both spin channels. The dashed lines indicate the Fermi energy 
of the left and right electrode. (b) displays the same as (a) for a bias voltage $V=-0.3$V.}
\label{fig4}
\end{figure*}

\begin{figure*}
\includegraphics[width=0.8\textwidth]{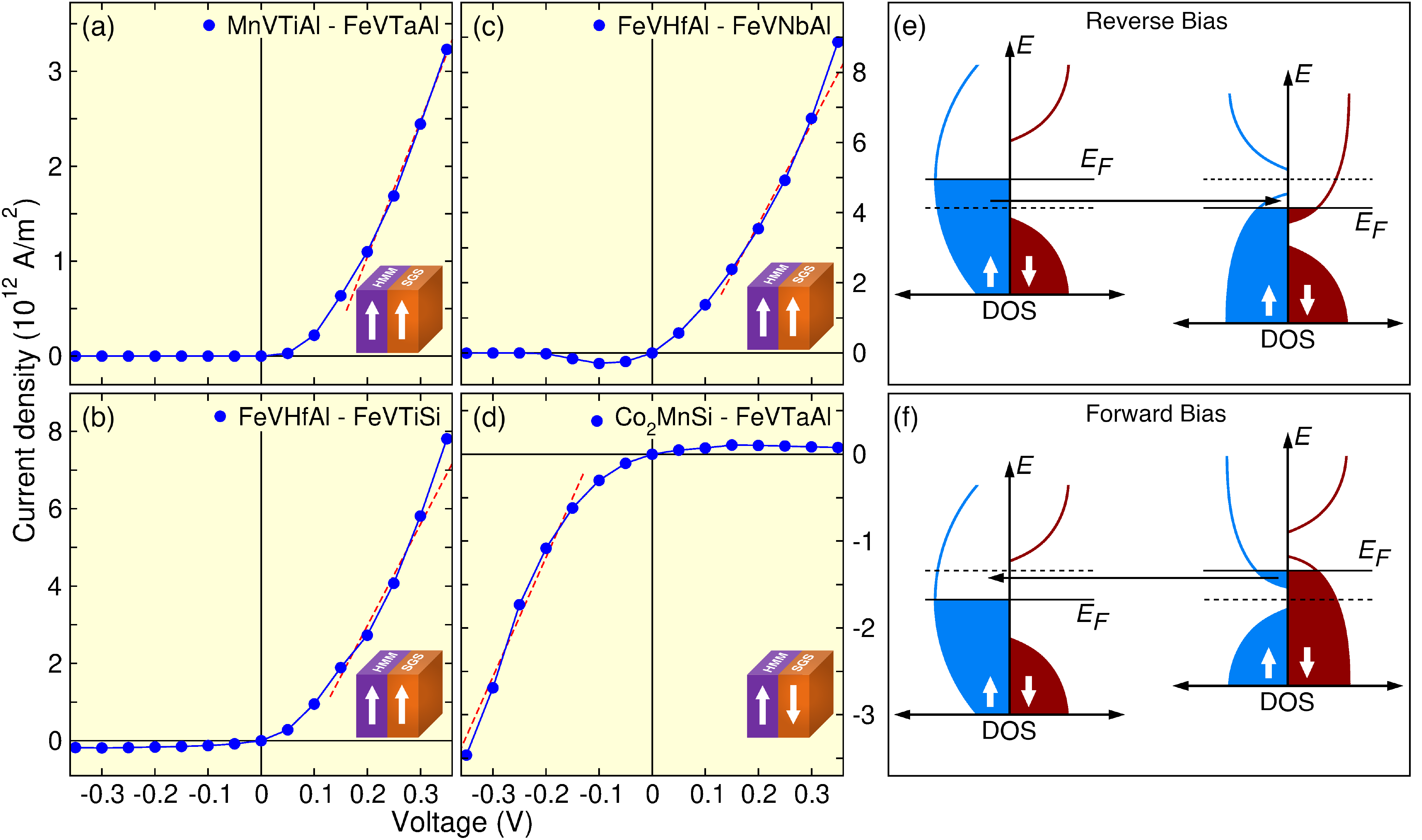}
\caption{(Color online) Calculated current-voltage ($I-V$) characteristics for all HMM-SGS 
junctions (a-d). The red lines display a linear fit. The coupling of the electrodes 
is displayed by a small image in the lower right corner. (e) and (f) show the origin of the
leakage current under reverse and forward bias for ferromagnetically and antiferromagnetically 
coupled HMM and SGS electrodes, respectively.}
    \label{fig5}
\end{figure*}

Up to now, we discussed the properties of the OSD for zero bias and will now focus on the current-voltage  
($I-V$) characteristics when a bias voltage is applied. Therefore, Fig.~\hyperref[fig4]{4\,(a)} and
\hyperref[fig4]{4\,(b)} illustrate the DDOS for both spin channels for the MnVTiAl$-$FeVTaAl junction 
under a bias voltage of +0.3\,V and -0.3\,V, respectively. Also the corresponding transmission spectra 
are presented there. For both, forward and reverse bias, the electronic and magnetic properties of both 
materials are not influenced by the bias voltage. The $I-V$ characteristics of the MnVTiAl$-$FeVHfAl 
junction presented in Fig.~\hyperref[fig5]{5\,(a)} can be explained on the basis of the DDOS. Under a 
forward  bias voltage majority spin electrons from the occupied states below the Fermi level in FeVTaAl 
can be transmitted to unoccupied states above the Fermi level in MnVTiAl. Thus, the transmission coefficient 
has a finite value in this case. However, for the minority spin electrons the transmission coefficient is 
zero because both materials have no states that could contribute to transport in the given voltage window. 
Thus, the forward current (on-current) is 100\% spin polarized. Also under an applied reverse bias voltage 
the transmission coefficient for minority-spin electrons is zero due to the energy gap in both materials.
On the other hand, the overlap of conduction and valence bands of opposite spin channels around the Fermi
energy in FeVTaAl gives rise to a non-zero transmission coefficient for 
majority-spin electrons, which leads to a leakage current that will be  discussed in detail in the following paragraph.

Before we discuss the origin of the leakage current, we will briefly comment on the  $I-V$ characteristics
of the other three OSDs, which are also presented in \cref{fig5}. As seen there, for all OSDs we obtain a linear  
behavior starting from around +0.15\,V for the ferromagnetic OSDs (MnVTiAl$-$FeVTaAl, FeVHfAl$-$FeVTiSi and FeVHfAl$-$FeVNbAl 
junctions) and around -0.15\,V in the case of the antiferromagnetic OSD Co$_2$MnSi$-$FeVTaAl. For the three ferromagnetic OSDs, the $I-V$ curves are more or less similar to each other, while 
the Co$_2$MnSi$-$FeVTaAl junction 
is in the off state for a forward bias and in the on state for a reverse bias, which 
is somewhat similar to the backward  diode~\cite{sze2007physics,murali2018gate,roy2015dual}. An interesting
feature of this latter OSD would be its dynamical configuration since the  magnetic coupling strength of the electrodes
at the interface is rather weak ($\sim $17 meV). Thus, by applying an external magnetic field, the $I-V$ 
curves of the diode can be reversed similar to the case of reconfigurable magnetic tunnel diode concept
in Ref.~\onlinecite{sasioglu2019proposal}. Returning back to the discussion of the $I-V$ characteristics, all four OSDs exhibit exactly zero 
threshold voltage $V_T$ under forward bias. It is worth to note that all semiconductor diodes have sizeable threshold voltages V$_T$ (V$_T$ $\sim$ 0.7 V for silicone $p-n$ diodes), which gives rise to the power dissipation ($P = V_T \cdot I$) in form of heat and thus this is an undesirable effect. The larger the value of the threshold voltage V$_T$, the higher is the power dissipation in a diode.
Furthermore, for all OSDs the leakage currents are small
compared to the on-currents.
The leakage current 
can be traced back to the small overlap of conduction and valence  band edges of opposite spin channels around the
Fermi level in the SGS electrode as schematically shown in Figs.\,\hyperref[fig5]{5\,(e)} and \hyperref[fig5]{5\,(f)} (see Refs.~\onlinecite{gao2019high} and \onlinecite{aull2019ab} for the band structure and DOS of the SGS materials). Band overlaps allow in the ferromagnetic
(antiferromagnetic) OSD the flow of majority spin electrons  from the occupied states of the HMM (SGS) material into 
the unoccupied states of the SGS (HMM) electrode.
Thus, this process gives rise to a small leakage current. However, the leakage current is absent in  the minority-spin channel in both cases due to the energy gaps in the electrode materials. Therefore, leakage currents can be prevented by using ideal SGS materials, i.e., without an overlap in conduction and valence band edges of opposite spin channels around the Fermi energy. 
Since FeVTiSi exhibits an overlap of 150\,meV while FeVNbAl and FeVTaAl possess overlaps of 45\,meV and 60\,meV, respectively, the FeVHfAl$-$FeVTiSi junction shows the largest leakage current.

Finally, we would like to comment on the on/off ratios and current densities of the OSDs.
The $I_{\rm{ON}}/I_{\rm{OFF}}$ ratios at $\pm \, 0.3$\,V vary between $30$ (FeVHfAl$-$FeVTiSi) and 
$10^5$ (MnVTiAl$-$FeVTaAl) at zero temperature. Since FeVTiSi possesses the largest overlap between the conduction and valence band edges
of opposite spin channels around $E_F$, 
the FeVHfAl$-$FeVTiSi junction shows the largest leakage current as discussed above and thus the 
lowest on/off ratio. From that point of view, also the on/off current ratios can be increased by 
using materials with ideal spin-gapless semiconducting behavior. Here, the MnVTiAl$-$FeVTaAl 
junction seems to be the best candidate for realizing the OSD. Another aspect that influences the 
on/off ratios is temperature, which is neglected in our transport calculations due to the technical 
limitation of the \textsc{QuantumATK}  package for spintronic materials as discussed in Ref.~\onlinecite{SpinDiode}. 
Temperature effects as well as the spin-flip excitations can further reduce the on/off current ratio in 
OSDs (see Ref.~\onlinecite{SpinDiode} for a detailed discussion). As for the current densities, 
the calculated values are comparable to the elementary metals and much higher than 
conventional $p-n$ or $p-i-n$  diodes~\cite{harris1980current,gibson1980recent,szydlo1982high,deng2003absence}.
It is worth to note that in transport calculations within the \textsc{QuantumATK} package
all inelastic 
scattering processes stemming from phonons as well as electrons and magnons are neglected. All these
neglected processes can substantially reduce the current density of the OSDs.

In conclusion, the OSD is a recently proposed concept in spintronics and requires materials with 
unique electronic properties, in particular half-metallic and spin-gapless semiconducting 
behavior. Since both properties have already been identified in the family of ordered 
quaternary Heusler compounds, this family is a preferable choice for the realization 
of such devices. Moreover, most of the compounds within this family possess very high 
Curie temperatures making them potential candidates for spintronic applications at 
room temperature. By using first principles DFT calculations combined with the NEGF
method, we proposed four different HMM-SGS junctions (or OSDs) within the family of 
quaternary Heusler compounds. All four OSDs show linear $I-V$ characteristics with 
zero threshold voltage $V_T$ in the on state and small leakage currents in the off 
state which can be attributed to the small overlap of conduction and valence band edges
of opposite spin channels around the Fermi level in the SGS electrode. In three of the 
designed OSDs, the HMM and SGS electrodes couple ferromagnetically, while in the 
Co$_2$MnSi$-$FeVTaAl junction this coupling is antiferromagnetic and thus this
diode can be configured dynamically via an external magnetic field. Furthermore, the 
zero threshold voltage $V_T$ is important for reducing the power consumption 
in a diode as it scales linearly with $V_T$. We hope that our 
results pave the way for the experimental fabrication of OSDs based on quaternary 
Heusler compounds.

\acknowledgments

\noindent This work was supported by the European Union (EFRE), Grant No: ZS/2016/06/79307 and by Deutsche Forschungsgemeinschaft (DFG) SFB CRC/TRR 227.

\section*{Data availability}

\noindent The data that support the findings of this study are available from the corresponding author upon reasonable request.

\bibliographystyle{aipnum4-2}
\bibliography{bibliography}

\end{document}